\documentclass[aps,twocolumn,prl,floatfix,showpacs]{revtex4}
\usepackage{eurosym}
\usepackage{amssymb}
\usepackage{amsmath}
\usepackage{graphicx}
\usepackage{bm}
\usepackage{xcolor}
 \newcommand{\bk}{{\bf k}}
\begin{document}

\title{Topological quantum phase transition between Fermi liquid phases in an Anderson impurity model}

\author{G. G. Blesio}
\affiliation{Instituto de F\'{\i}sica Rosario (CONICET) and Universidad Nacional de Rosario, 
Bv. 27 de Febrero 210 bis, 2000 Rosario, Argentina}

\author{L. O. Manuel}
\affiliation{Instituto de F\'{\i}sica Rosario (CONICET) and Universidad Nacional de Rosario, 
Bv. 27 de Febrero 210 bis, 2000 Rosario, Argentina}

\author{P. Roura-Bas}
\affiliation{Centro At\'{o}mico Bariloche and Instituto Balseiro, Comisi\'{o}n Nacional
de Energ\'{\i}a At\'{o}mica, CONICET, 8400 Bariloche, Argentina}

\author{A. A. Aligia}
\affiliation{Centro At\'{o}mico Bariloche and Instituto Balseiro, Comisi\'{o}n Nacional
de Energ\'{\i}a At\'{o}mica, CONICET, 8400 Bariloche, Argentina}

\begin{abstract}
We study a generalized Anderson model that mixes two localized configurations 
--one formed by two degenerate doublets and the other by a triplet with single-ion 
anisotropy $DS_z^2$-- by means of two degenerate conduction channels.
The model has been derived for a single Ni impurity embedded into an O-doped Au chain. 
Using the numerical renormalization group, we find a topological quantum phase transition, 
at a finite value $D_c,$ between two regular Fermi liquid phases of high (low) conductance and 
topological number $2 I_L/\pi  = 0$ (-1)  for $D < D_c$ ($D > D_c$), where $I_L$ is the 
well-known Luttinger integral.
At finite temperature the two phases are separated by a non-Fermi liquid phase with fractional 
impurity entropy $\frac{1}{2}{\rm ln}2$ and other properties which remind those of the two-channel Kondo model.
\end{abstract}

\maketitle


\textit{Introduction-} Quantum phase transitions (QPTs) observed in transport through 
molecular systems in which two electrons play a relevant role have been a subject of 
interest recently \cite{roch08,parks10,florens11}. In general, in nanoscopic systems with 
more than one electron, the spin-orbit coupling is important and leads to the single-ion
anisotropy $DS_z^2$, where $S_z$ is the total spin of the molecule or quantum dot 
\cite{parks10,florens11,katsaros10,jespersen11,hiraoka17}. The relative magnitude of $D$ can be tuned 
experimentally \cite{parks10,oberg13,heinrich15,hiraoka17}. On the other hand, experiments
with mechanically controllable break junctions made possible to create one-dimensional atomic chains 
of several elements, and measure the conductance through them \cite{ohnishi98,rodrigues03}.  

Some QPTs are topological QPTs (TQPTs): even if some other properties vary continuously at the 
transition, a topological integer (related with a geometrical Berry phase or the topology of each 
thermodynamic phase) jumps at the TQPT. Examples of this kind of transitions 
are several charge and spin TQPTs observed in one-dimensional models in which the nearest-neighbor 
hopping depends on the occupation \cite{aligia00,aligia07} as in cold-atom lattices \cite{duan08}, or the 
Hubbard model with alternative on-site energies \cite{fabrizio99,torio01}, for which the topological 
transition might be observable in transport through arrays of quantum dots or molecules \cite{aligia04}.

On the other hand, in condensed matter physics, the Luttinger theorem \cite{luttinger60,luttinger60b}, which states 
that the volume of the Fermi surface is determined by the particle density and remains unaltered 
by interactions, and Friedel sum rules \cite{langreth66,yoshimori82}, which relate the occupancy of impurity 
states with the corresponding spectral density at the Fermi energy, have been crucial for our present 
understanding of many interacting systems that behave as Fermi liquids at zero temperature. 
The so-called Luttinger integral $I_L$ \cite{note} enters the demonstrations in these works and 
it was generally assumed to vanish. However, recently a group of researchers found that $I_L$ can take
three different values in an impurity model in phases with regular low-energy Fermi liquid behavior \cite{curtin18,nishikawa17}.
This is surprising since only $I_L=0$ was expected in a regular Fermi liquid, according to its perturbative calculation 
in the seminal work by Luttinger and Ward \cite{luttinger60b}, where the Fermi liquid was considered 
adiabatically connected to a system of non-interacting electrons. 
More recently, Seki and Yunoki \cite{seki17} showed that the Luttinger integral, which is the deviation 
of the Luttinger volume from the non-interacting limit, can be interpreted as a winding number of the ratio 
between the determinants of the non-interacting and interacting single-particle Green's functions. The combination of this topological 
interpretation of $I_L$ and the finding of its non-zero values \cite{curtin18,nishikawa17} opens the possibility 
of topologically non-trivial Fermi liquid phases, that are not adiabatically connected with non-interacting systems and, therefore,
they can be termed as {\it non-Landau} Fermi liquids.

In this Letter, we describe the TQPT that, driven by the single-ion anisotropy $D$, takes place in a 
simple impurity model which, in the Kondo limit, consists in a spin 1 screened by two conduction channels. 
This model has been derived from \textit{ab initio} calculations and describes transport through Ni atoms 
in O doped Au chains \cite{dinapoli15,barral17}. The oxygen doping has the effect of pushing up the $5d_{xz}$ 
and $5d_{yz}$ bands of Au (with $z$ along the chain direction), which are below the Fermi energy in 
pure Au chains \cite{barral17,thijssen06,dinapoli12}. 
The two conduction channels correspond to the degenerate orbitals of $xz$ and $yz$ symmetry, 
and they hybridize with the corresponding Ni $3d$ orbitals.
Solving the model by means of the numerical renormalization group (NRG), 
we find that the transition occurs at the finite critical anisotropy $D_c \approx 2.57\; T_K^0$, 
where $T_K^0$ is the Kondo temperature for $D=0$. This TQPT separates two regular Fermi liquid phases:  
for $D < D_c$ the impurity spin is Kondo screened, while for $ D > D_c$ it is  
quenched by the anisotropy. 
For $D < D_c,$ as the temperature $T \rightarrow 0,$ the electrical conductance is large and agrees with the usual Friedel sum rule 
with $I_L=0$, as in Fermi liquids adiabatically connected with a non-interacting system. 
Instead, for $D > D_c$ and $T=0$,  the conductance is small and satisfies a generalized Friedel 
sum rule with $I_L=-\frac{\pi}{2},$ corresponding to a {\it non-Landau} Fermi liquid. 
Furthermore, for $D \approx D_c$ and in a finite interval of temperatures $T^{*}(D) \leq T \lesssim T^0_K,$ 
where $T^*(D) \rightarrow 0$ as $D \to D_c$ \cite{nishikawa12}, 
there is a critical quantum regime whose electrical transport and thermodynamics properties 
correspond to a non-Fermi liquid behavior. It is worth to mention that 
$T_K^0$ can be tuned by stretching
the gold chains, rendering it possible to observe the transition experimentally. 

\textit{Model-} 
We consider the Hamiltonian that describes a system containing a Ni atom in a substitutional position within a Au chain 
doped with a small amount of oxygen ($\sim 14\%$) \cite{dinapoli15,barral17}, describing charge fluctuations 
between $3d^8$ and $3d^9$ Ni configurations. 
It can be written as \cite{sm}
\begin{eqnarray} 
H&=&\sum_{M_{2}}(E_{2}\!+\!
D M_{2}^{2})|M_{2}\rangle \langle
M_{2}|+\sum_{\alpha M_{1}}E_1|\alpha M_{1}\rangle \langle \alpha M_{1}| + \nonumber\\
&+&\sum_{\nu k\alpha \sigma }\varepsilon _{\nu k}c_{\nu k\alpha \sigma
}^{\dagger }c_{\nu k\alpha \sigma }+
\label{Hamiltonian}
\\
&+&\sum_{\substack{ M_{1}M_{2} \\\alpha\nu k\sigma}}V_{\nu}
\langle 1 M_{2}  |\frac{1}{2}\frac{1}{2}M_{1}\sigma \rangle 
( | M_{2}\rangle \langle \alpha M_{1}|c_{\nu k\alpha \sigma }+\mathrm{H.c.} ),  \nonumber 
\end{eqnarray}
where $E_i$ and  $M_i$ indicate the energies and the spin projections along the chain, respectively,   
of states with $i=1, 2$ holes in the $3d$ shell of the Ni impurity; 
$|\alpha M_{1}\rangle$ is the state with one hole with symmetry $\alpha$ ($xz$, $yz$) and spin $M_1$.
$D$ is the Ni uniaxial magnetic anisotropy. 
The operator $c_{\nu k\alpha \sigma}^{\dagger }$ creates a hole 
with symmetry $\alpha$ and energy $\varepsilon_{\nu k}$ (relative to the Fermi level $\varepsilon_F=0$)
in the $5d$ shell of the Au atom,
where $\nu=L, R$ denotes the left or the right side of the Ni atom, 
respectively. $\langle 1 M_{2}  |\frac{1}{2}\frac{1}{2}M_{1}\sigma \rangle $ are
Clebsh-Gordan coefficients.
The hopping $V_{\nu}$ characterizes the tunneling between the Ni and Au states, and it
enters the hybridization function $\Delta = \pi \sum_{\nu k}|V_\nu|^2\delta(\omega-\varepsilon_{\nu k}),$
assumed independent of energy.

The ground-sate configuration of the Ni atom has two holes in the degenerate $3d_{xz}$, $3d_{yz}$  
orbitals coupled to spin $S=1$. The state $|M_{2}\rangle$ with $M_2=0$ is lower in energy
than those with $M_2= \pm 1$ by an energy that has been estimated in $D \approx 8.5$ meV
solving exactly the atomic model for the 3d$^8$ configuration including all interactions and spin-orbit 
coupling \cite{dinapoli15}. 

We solve the Hamiltonian (\ref{Hamiltonian}) by means of NRG, 
as implemented in the Ljubljana open source code \cite{zitko14}. We use a discretization parameter $\Lambda = 3$, 
and we keep up to 10000 states. The results are $z$-averaged with $N_z$ up to 4. 
In this work, we have chosen $\varepsilon_d \equiv E_1-E_2 = -0.02$ and $\Delta = 0.1$ 
in units of the conduction half-bandwidth $W$. For this ratio $\varepsilon_d/\Delta$, 
the system is in the Kondo regime, but close to the mixed valence regime (MVR),
and the different regimes that we want to display come out more clearly.
The corresponding Kondo temperature is $T^0_K \simeq 1.245 \times 10^{-3}$ for $D=0$, obtained through the 
usual condition $G(T^0_K)= \frac{1}{2}G(T\to 0)$, where $G(T)$ is the differential conductance $G=dI/dV$.
However, the occupancy found
in 
{\it ab-initio} calculations \cite{dinapoli15,barral17} indicates that the system is closer to the MVR 
and with a Kondo temperature near 6 meV, as explained in the Supplemental Material \cite{sm}.
The corresponding value  $D_c \simeq 2.57 T^0_K \sim$ 15 
meV roughly falls in the range of the estimated $D$ for Ni atoms in 
O doped Au chains, particularly taking into account that $D$ can be reduced by stretching or tuned by 
further doping the gold chains.

\textit{Generalized Friedel sum rule-}
Using conservation laws, the impurity spectral function per orbital and spin, at the Fermi level 
and $T=0,$ is given by \cite{yoshimori82}
\begin{equation}
A_{d\alpha\sigma}(\omega=0)=\frac{1}{\pi \Delta} \sin ^{2}(\delta_{\alpha\sigma}),
\label{fsr}
\end{equation}
where, taking into account explicitly the spin degeneracy, the phase shift is
\begin{equation}
\delta _{\alpha\sigma}= \frac{\pi}{2} \langle n_{d\alpha} \rangle -I_L.
\label{delta}
\end{equation}
$n_{d\alpha}=\sum_{M_2}|M_2><M_2|+\sum_{M_1}|\alpha M_1><\alpha M_1|$ is the hole
occupation number of the Ni $\alpha$ orbital, and the 
Luttinger integral $I_L$, which in our case is independent of orbital and spin indices, 
is defined as
\begin{equation}
I_L ={\rm Im} \int_{- \infty}^{0} d \omega G_{d\alpha \sigma}(\omega) 
\frac{\partial \Sigma_{d\alpha \sigma}(\omega)}{\partial \omega},  
\label{il}
\end{equation}
where $G_{d\alpha \sigma}(\omega)$ is the impurity Green function for orbital $\alpha$ and spin $\sigma$ and 
$\Sigma_{d\alpha \sigma}(\omega)$ is the corresponding self energy.

As explained above, $I_L$ vanishes for a Fermi liquid, when it is perturbatively calculated from a non-interacting electronic 
system \cite{luttinger60b}. However, recently \cite{curtin18,nishikawa17} it was found that this is not always the case for local 
Fermi liquids, 
while a topological interpretation of $I_L$ was provided for extended systems \cite{seki17}.

\textit{NRG results-} 
To localize the TQPT we use the differential conductance, which 
is easily accessible experimentally \cite{ohnishi98,rodrigues03}. 
The conductance per channel is given by 
\begin{equation}
G_{\alpha}(T)=G_0 \sum_\sigma \frac{\pi \Delta}{2}\int d\omega \left( -\frac{\partial f(\omega )}{\partial \omega }\right) 
A _{d\alpha \sigma }(\omega )
\label{gas}, 
\end{equation}
where $f(\omega)$ is the Fermi function and $G_0=2e^2/h$ is the quantum of conductance.
Using Eqs. (\ref{fsr}) and (\ref{delta}) we have, at zero temperature, the generalized Friedel sum rule for the conductance
\begin{equation}
G_{\alpha}(0)=G_0 \sin ^{2}\left(\frac{\pi}{2} \langle n_{d\alpha} \rangle -I_L\right).  
\label{gt0}
\end{equation}

\begin{figure}[ht]
\begin{center}
\includegraphics*[width=0.85\columnwidth]{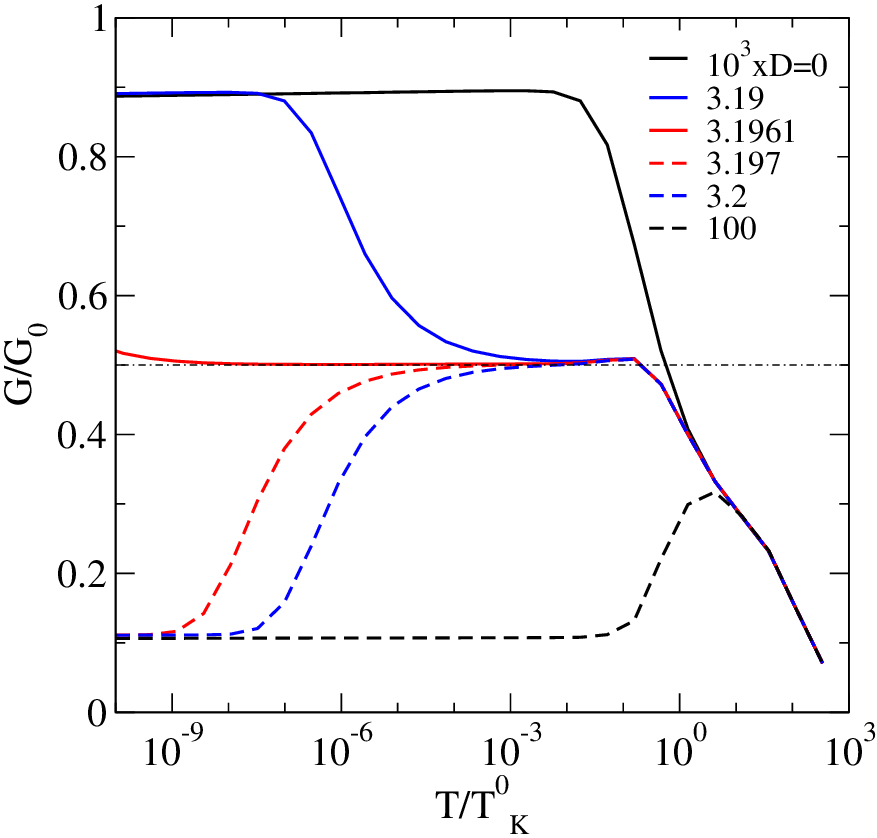}
\caption{(Color online) Electrical differential conductance as a function of $T/T^0_K$ for several values of the single-ion anisotropy $D.$}
\label{conductance}
\end{center}
\end{figure}

In Fig. \ref{conductance}, the conductance per channel as a function of $T/T_K^0,$ for several positive 
values of the single-ion anisotropy $D$ is shown. $G_\alpha$ has an abrupt change as $D$ is varied across its critical value 
$D_c \simeq 0.003196,$ and two regimes are easily characterized according to the behavior of $G_\alpha$ at the 
lowest temperatures. 
For $D < D_c,$  $G_\alpha(T\!\to\!0)$ takes a large value, which corresponds to the Friedel sum rule 
(\ref{gt0}) with $I_L=0$. In this case, for low temperatures we expect a fully Kondo screened impurity, 
leading to the usual Fermi liquid phase. Note that, as we are working far away from the particle-hole symmetric point 
(only $3d^8$ and $3d^9$ Ni configurations are considered), the occupation number per impurity orbital is less than 1 
($\langle n_{d\alpha}\rangle \simeq 0.788$, almost constant with varying $D$) and $G_\alpha$ does not reach the unitary limit. 
It can be seen that, for small $D$, the conductance exhibits fingerprints of the magnetic anisotropy for temperatures of 
the order of $D$, like the shoulder that develops before it goes to its Kondo limit with decreasing $T$. 

On the other hand, for $D > D_c$, $G_\alpha (T\to 0)$ goes to a low value, 
corresponding to $I_L = -\frac{\pi}{2}$ in (\ref{gt0}). This case corresponds to the impurity spin quenched by $D$, as its 
ground state has spin projection $S_z=0$. Again, we expect a Fermi liquid at low $T$, but now this phase yields the non-trivial 
Luttinger integral $I_L = -\frac{\pi}{2}$. For $D \gg D_c,$ $G_\alpha$ takes small values for any temperature. 

For $D$ close to $D_c$, below and above, the conductance has a clear plateau at the precise value $G_0/2$, 
characteristic of the two-channel Kondo (2CK) effect. 

Around $D_c$  we define a characteristic energy for each phase. For $D < D_c,$ 
$T_K^*(D)$ is computed through the condition $G_\alpha(T_K^*(D))=(G_\alpha(T\!\to\!0)-0.5 G_0)/2,$ corresponding to 
the onset of the fully Kondo screening of the impurity, while for $D > D_c$, we take $T^*_q(D)$ which satisfies 
$G_\alpha(T^*_q(D)) = (0.5 G_0-G_\alpha(T\!\to\! 0))/2,$ and it signals the onset of the impurity spin quenching. 
We find that, as it corresponds to a quantum critical point \cite{nishikawa12},
these energies vanish as $D \to D_c$. Surprisingly for a Kondo screening energy scale, 
$T^*_K(D)$ has a potential law dependence on $D$: $$T^*_K(D) \propto T_K^0 \left(\frac{D_c-D}{D_c}\right)^2.$$ 
On the other hand, $$T^*_q(D) \propto T_K^0 \exp \left[-c\left(\frac{T_K^0}{D-D_c}\right)^{1/4}\right],$$  where $c$ is a constant of order of one; 
a similar result was found in the $S=1$ underscreened Kondo model \cite{cornaglia11}.

\begin{figure}[ht]
\begin{center}
\includegraphics[width=0.85\columnwidth]{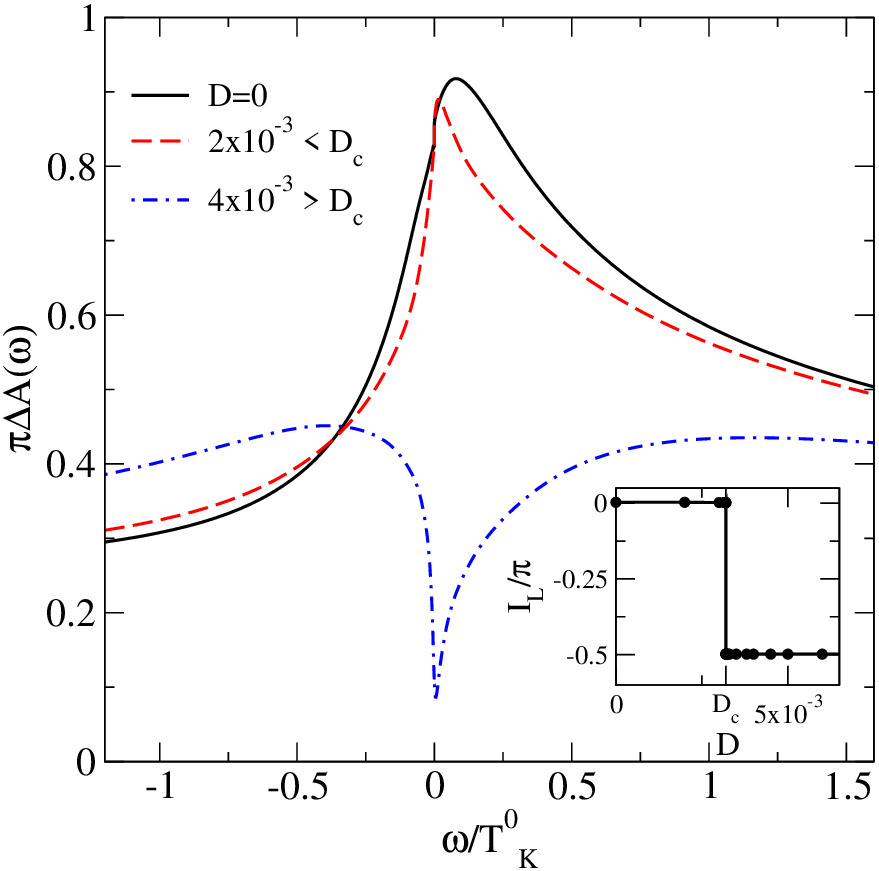} 
\caption{(Color online) Impurity spectral function $A_{d\alpha \sigma}(\omega)$ for three different anisotropies $D$, at 
$T \simeq 10^{-6}T_K^0$. Inset: Luttinger integral $I_L$ as a function of $D$.}
\label{spectral}
\end{center}
\end{figure}

The impurity spectral function $A_{d\alpha\sigma}(\omega)$ near the Fermi level is presented in the main panel of 
Fig. \ref{spectral}, for three different values of $D,$ at the very low temperature $T \simeq 10^{-6}T_K^0.$ 
A Kondo resonance is clearly visible for all $D < D_c.$ As $D$ increases, the Kondo peak moves towards the Fermi level and its width decreases.
At $D_c$ the resonance abruptly disappears, and it is replaced by a narrow dip just at $\omega=0$. In the supplemental material \cite{sm}, 
we show that for the related $S=1$ Kondo impurity model with two conduction channels, very close to its critical point, 
the spectral function (defined through the $t-$matrix) takes half of its Kondo-screened value, being this another hallmark of the 2CK. 

With the NRG technique it is not an easy task to obtain reliable values of $I_L$ by computing it directly from Eq. (\ref{il}), due to numerical inaccuracies 
in the self-energy evaluation  \cite{curtin18}. Instead, we calculate $I_L$ through the generalized Friedel sum rule for the conductance (\ref{gt0}). 
The obtained $I_L$ is displayed as a function of $D$ in the inset of Fig. \ref{spectral}. 
It can be seen that $I_L$ takes only two discrete values: $I_L =0, -\pi/2$, with an abrupt jump at $D_c$. 
This is not fortuitous as $I_L$ is closely related with the winding number of the ratio 
$D_d(z) = G^{0}_{d\alpha\sigma}(z)/G_{d\alpha\sigma}(z)$ between the non-interacting and interacting impurity Green's functions, 
around the origin in the complex plane $D_d$ (see supplemental material \cite{sm}):
\begin{equation}
 I_L = \pi \lim_{T\to 0}\oint_\Gamma \frac{dz}{2\pi i} n_F(z) \frac{\partial \ln D_d(z)}{\partial z},
\end{equation}
where the contour $\Gamma$ encloses the real axis. So, the two Fermi liquids for $D \lessgtr D_c$ can be topologically distinguished by $I_L$, being topologically 
trivial (non-trivial) synonymous of adiabatically (non-adiabatically) connected to a non-interacting system. As a consequence, the quantum critical transition between the 
two Fermi liquids (a {\it Landau}- and a {\it non-Landau} Fermi liquid) at 
$D_c$ has a topological character. With the necessary caution due to the difficulties of the NRG computation of the self-energy \cite{nota_zitko}, 
we have checked that its imaginary part ${\rm Im}\;\Sigma_{d\alpha \sigma}(\omega)$ behaves quadratically as a function of frequency close to the Fermi level, 
for both Fermi liquids $I_L = 0, -\pi/2.$  However, for $D \simeq D_c$ a singularity appears just on the Fermi level, being responsible of the non-trivial $I_L$.
We conjeture that this singularity is related with the simultaneous creation, as $D \to D_c$ from below, 
of a zero and a pole of the impurity Green's function at the Fermi level, as it happens 
in an analogous way in other topological transitions in extended systems \cite{sakai09,gurarie11}. 
\begin{figure}[ht]
\begin{center}
\includegraphics[width=0.75\columnwidth]{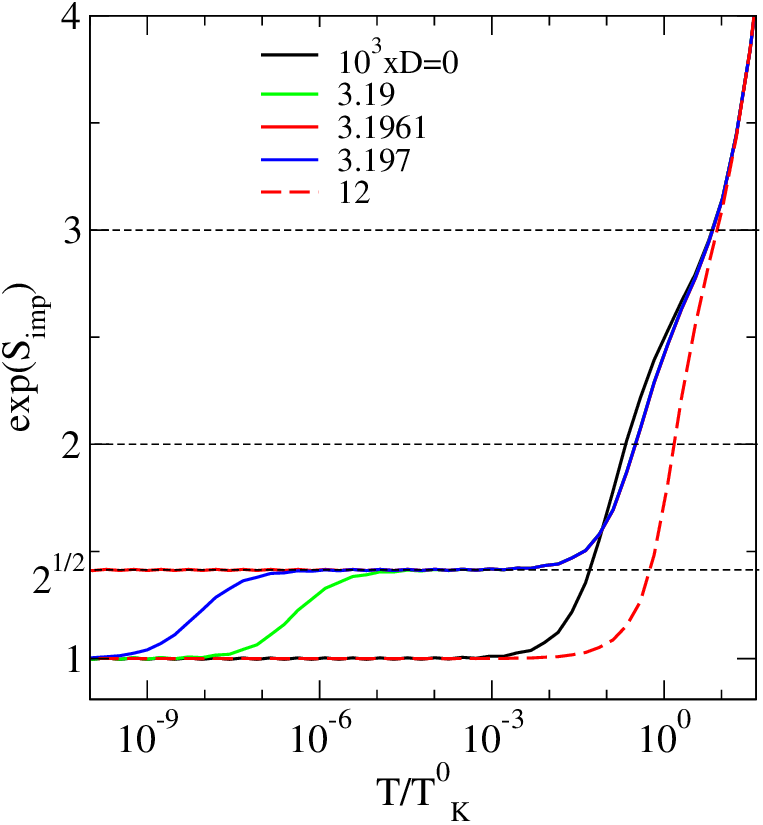} 
\caption{(Color online) Impurity entropy as a function of temperature for several single-ion anisotropy $D$ close to 
the 
TQPT.}
\label{entropy}
\end{center}
\end{figure}

In order to further characterize the critical region, the impurity contribution to the entropy as a function of temperature is plotted in 
Fig. \ref{entropy}. It can be clearly seen that, for $D$ close to $D_c,$ there is a plateau at $S_{\rm imp} = \frac{1}{2}\ln(2)$, the fractional 
entropy usually associated with the 2CK physics \cite{mitchell12}. Also, a shoulder at $S \simeq 3$ is noticeable, corresponding to the three-fold 
degeneracy of the $S=1$ impurity states at intermediate temperature. For other parameters (not shown in this work), 
this shoulder 
transforms in a clear plateau. At higher temperatures, out of the figure, there is a plateau 
at $S_{\rm imp}=7$ corresponding to the total number of localized impurity states in the model.

Another signature of 2CK-like behavior close to $D_c$ is the fact that the NRG spectrum, as a function of the NRG 
iteration number $N$ (see supplemental material \cite{sm}), has an extended plateau for intermediate $N,$ corresponding to a (unstable) fixed-point 
without the typical odd/even alternation and uniform level spacing of the conventional Kondo effect \cite{pang91}.

\begin{figure}[ht]
\begin{center}
\includegraphics*[width=0.6\textwidth]{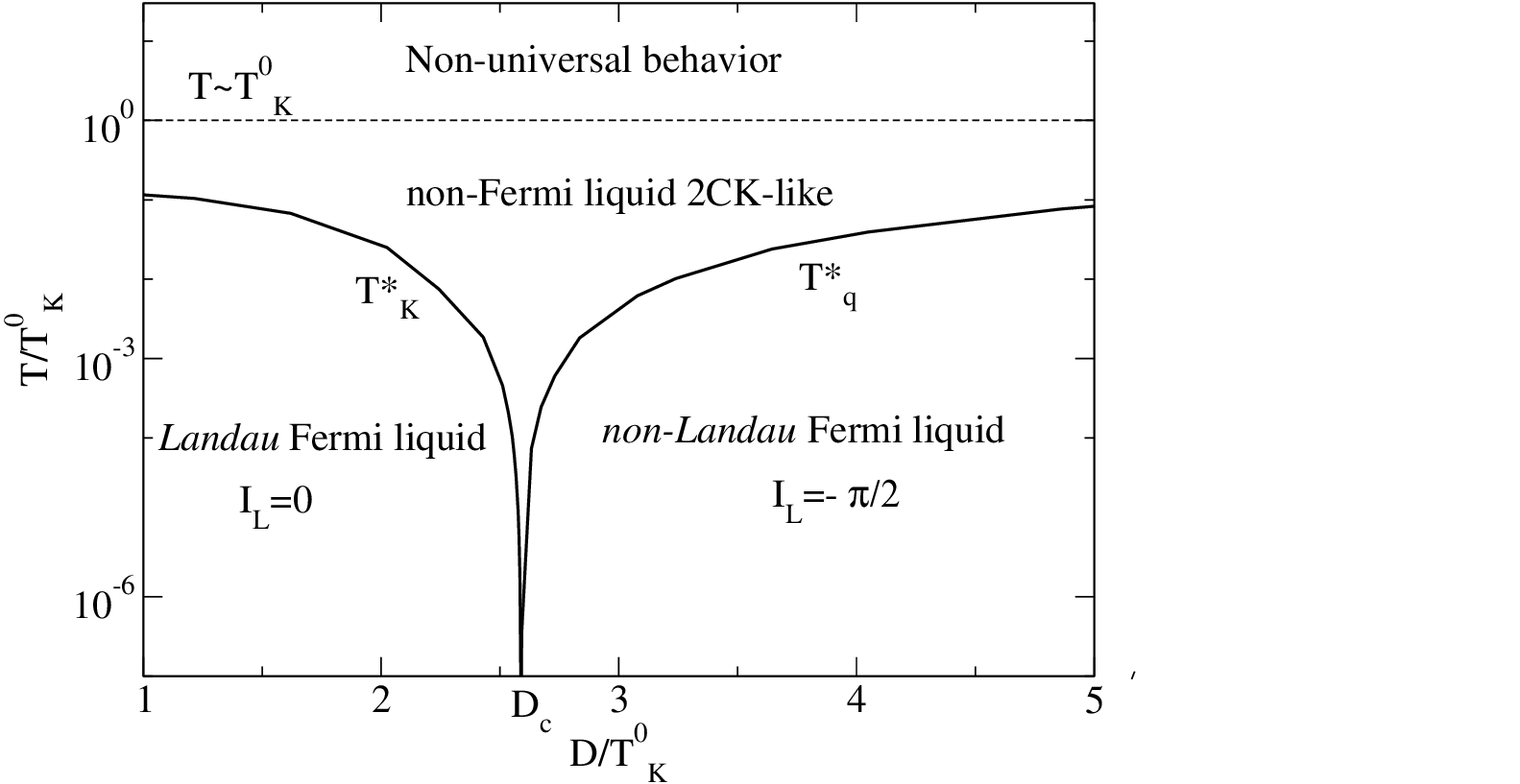}
\caption{Sketch of the phase diagram of the $S=1$ impurity model as a function of temperature and anisotropy. The solid lines indicate crossover regions between 
Fermi and non-Fermi liquid behaviors, while the dashed one signals the onset of non-universal behavior at higher temperatures.}
\label{phasediagram}
\end{center}
\end{figure}
 
Using the energy scales $T_K^*$ and $T^*_q$, we can summarize our findings in the phase diagram sketched in Fig. \ref{phasediagram}, with 
its ``classical'' Fermi liquid regions at both sides of the critical point, while the usual (non-Fermi liquid) quantum critical wedge emerges 
from the quantum critical point $D_c$ at zero temperature. 

\textit{Summary-} We have found a topological quantum phase transition between two Fermi liquids through an 
intermediate non-Fermi liquid 2CK-like phase in a simple model consisting in a $S=1$ impurity coupled to two conduction bands, 
where the driving parameter of the transition is the single-ion magnetic anisotropy $D.$ 
This model has experimental relevance for transport through nanostructures formed by Ni impurities in O-doped gold chains. 
The relative magnitude of anisotropy to the Kondo temperature $T_K^0$ can be experimentally tunable, rendering possible to observe the 
transition: $D/T^0_K$ can be modified by changing the effective $\varepsilon_d$ of the Ni atom by doping or it can be increased 
by mechanically streching the gold chains \cite{parks10}.
We expect that our work will stimulate further experimental work in similar systems. 
In particular, it would be interesting to find experimental probes that can distinguish between Fermi liquids characterized by different
values of the topological invariant Luttinger integral.

\textit{Acknowledgements-} We acknowledge financial support provided by PIP 112-201501-00506 of CONICET
and PICT 2013-1045 of the ANPCyT.


 \widetext
 \pagebreak
 \setcounter{equation}{0}
 \setcounter{figure}{0}
 \setcounter{page}{1}
 \makeatletter
 \renewcommand{\theequation}{S\arabic{equation}}
 \renewcommand{\thefigure}{S\arabic{figure}}
 \renewcommand{\bibnumfmt}[1]{[S#1]}
 \renewcommand{\citenumfont}[1]{S#1}

\begin{center}
 {\bf \large Supplemental material for: Topological quantum phase transition between Fermi liquid phases in an Anderson impurity model}
\end{center}

We begin expressing the impurity Hamiltonian in terms of fermionic operators, by means of a 
two-orbital Anderson impurity model, in order to be solved with the ``NRG Ljubljana'' package.
In Section II, we discuss the estimation of realistic parameters from {\it ab-initio} calculations 
for the Anderson impurity model. In Section III, we discuss some aspects of the NRG spectra.
In Section IV, we present results for the $S=1$ Kondo impurity model with single-ion anisotropy $D$ and coupled to
 two degenerate conduction bands, that show that topological quantum phase transitions at finite $D_c,$ 
 as we have found in  the Anderson  model, seem to be generic for this kind of fully screened anisotropic 
 $S=1$ impurity  systems. 
 Finally, in Section V, we give a topological interpretation of the Luttinger integral. 

\vskip 1cm

\section{Two-orbital Anderson impurity model used in NRG calculations} 
In order to use the ``NRG Ljubljana'' code \cite{zitko14-sm} for the resolution of Hamiltonian (1) --whose impurity degrees of 
freedom are expressed in terms of Hubbard operators--, we resort to a degenerate two-orbital 
Anderson impurity model, which involves only fermionic operators, and therefore it can be treated 
straightfowardly with the above mentioned code. 
The Anderson Hamiltonian is given by 
\begin{eqnarray} \label{ham-sm}
H & = & \sum_{\alpha \sigma} \varepsilon_{d} d^\dagger_{\alpha\sigma} d_{\alpha\sigma} + 
U\sum_{\alpha}n_{d\alpha\uparrow}n_{d\alpha\downarrow} + U' n_{d_{xz}}n_{d_{yz}}-
\nonumber \\
 &  - & J_H {\vec S}_{d_{xz}} \cdot {\vec S}_{d_{yz}} + 
+J'\left(d^\dagger_{xz\uparrow}d^{\dagger}_{xz\downarrow}d_{yz\downarrow}d_{yz\uparrow}+{\rm H.c.}\right) +
D S_z^2 +\label{and-sm} \\
& + & \sum_{\nu k\alpha \sigma }\varepsilon _{k}c_{\nu k\alpha \sigma }^{\dagger }c_{\nu k\alpha \sigma }+ 
\sum_{\nu k \alpha \sigma}\left( V_{\nu} {c}^\dagger_{\nu k \alpha\sigma}{d}_{\alpha \sigma} + {\rm H.c.}\right),
\nonumber 
\end{eqnarray}
where $d^\dagger_{\alpha \sigma}$ creates a hole with energy $\varepsilon_{d}$ in the $3d$ shell of the 
Ni atom with symmetry $\alpha= xz, yz$, 
while $c_{\nu k\alpha \sigma}^{\dagger }$ creates a hole in the $5d_{\alpha}$ shell of the Au atom and 
$\nu=L, R$ denotes the left or the right side of the Ni atom, respectively.
The energies $\varepsilon_d$, $\varepsilon_k$ are measured relative to the Fermi level, $\varepsilon_F = 0$.
$U$ ($U'$) is the intra- (inter-) orbital Coulomb repulsion, $J_H$ the ferromagnetic Hund exchange, $J'$ the pair-hopping 
parameter, and $D$ the 
uniaxial magnetic anisotropy. 
The hopping $V_{\nu}$ characterizes the tunneling between the Ni and Au states at each side, and it
enters the hybridization $\Delta=\pi \sum_{\nu k}|V_{\nu}|^{2}\delta (\omega -\varepsilon_{k}),$ 
assumed independent of energy.

To obtain the correct ground state configuration of the Ni atom --consisting of two holes in the degenerate 
3d$_{xz}$, 3d$_{yz}$ orbitals Hund-coupled to spin $S=1$, but close to the mixed valence regime due to 
the fluctuations between 3d$^8$ and 3d$^9$ configurations \cite{dinapoli15-sm,barral17-sm}--, we have to project out some states.  
The configurations with an electron occupancy equals to 0, 3, and 4 are removed exactly in the used 
Hilbert space, while the singlet states with 
2 electrons are projected out assigning them a very high energy. Thus, we take $U'= J_H/4$ and $U=J'=J_H$, with $J_H$ large enough 
($J_H \sim 1000W$, $W=1$ as the half-bandwidth of the conduction electrons) 
so that triplet and one-hole states remain the only relevant (low-energy) states, separated between them by an energy $\varepsilon_d$.

\section{Anderson impurity model parameters for Ni atoms in O-doped Au chains}

LDA calculations give us access to some parameters of the Anderson Hamiltonian \cite{dinapoli15-sm}: conduction electron 
half-bandwidth, hybridization, and on-site energy $\varepsilon_d.$
Also, LDA yields the hole occupancy $n_d$ of the 3d orbitals. Of all these magnitudes, $\varepsilon_d$ is by far the least 
reliable quantity because, to some extent, the {\it ab initio} computed orbital energies include 
contributions from Coulomb repulsions at a mean-field level, that should not be present in the bare 
$\varepsilon_d$ of the Anderson model. Instead of taking the LDA $\varepsilon_d$ as the bare orbital
energy, it is better to fix it by asking that the Anderson model and LDA occupancy numbers coincide \cite{nunez13-sm}.

For Ni atoms in O doped Au chains, LDA estimations are $W=5$ eV, $\Delta = 0.115$ eV, and $\varepsilon_d=-2\Delta=-0.23$ eV \cite{dinapoli15-sm}. 
If we consider these values in the Anderson model, we obtain a NRG Kondo temperature that 
is very low, $T^0_K \simeq 0.1$ meV ($D_c \simeq 0.45$ meV). This is understandable as the hole occupancy 
number, $n^{NRG}_d=1.943,$ indicates that, for this on-site energy, the impurity is well inside of the Kondo 
regime (this implies a 
low $T_K$), while, on the other hand, the hole occupancy obtained with LDA 
($n_d=1.44$, taken from Table I, first row of Ref. \onlinecite{dinapoli15-sm}) 
corresponds to a regime 
close to mixed valence. So, this a clear indication that the LDA estimated $\varepsilon_d$ is not a good input for the Anderson 
impurity Hamiltonian, as its value is well below the Fermi level. 
 Furthermore, the Haldane shift of $\varepsilon_d,$ in the $S=1$ case, is given by \cite{blesio18-sm}
 \begin{equation}
  \varepsilon_d^* = \varepsilon_d - \frac{\Delta}{2\pi}\log \frac{W}{\Delta},
 \end{equation} 
 and it pushes the bare on-site energy to even lower values.

To get the same hole occupancy as in the LDA calculations, we need to take a positive $\varepsilon_d \simeq 0.93 \Delta,$ leaving 
the other parameters fixed, in the Anderson impurity model. 
For this $\varepsilon_d$, the Kondo temperature is $T_K^0 = 6$ meV, so the critical anisotropy $D_c \sim 15$ meV, above 
the estimated $D$ ($\simeq 8$ meV)  of Ni atoms in O-doped gold chains. Thus, it is possible to promote the topological quantum phase transition 
reducing the Kondo temperature $T_K^0$ by means of the mechanical stretching of the gold chains. 

In our work, in order to highlight the non-Fermi liquid region, 
we consider an intermediate relative value of the on-site energy, $\varepsilon_d = -\Delta/5$.  The corresponding 
hole occupancy, $n_d = 1.576,$ is a bit above the value given by LDA, however the impurity is in the crossover region between Kondo and 
mixed valence regimes, as infered from LDA results for the occupancies.
For any parameter set choice that corresponds to the Kondo regime or that crossover region, the qualitative impurity behavior remains the same.

\section{NRG energy spectrum for the Anderson impurity model}
\begin{figure}[ht]
\begin{center}
\includegraphics*[width=0.5\columnwidth]{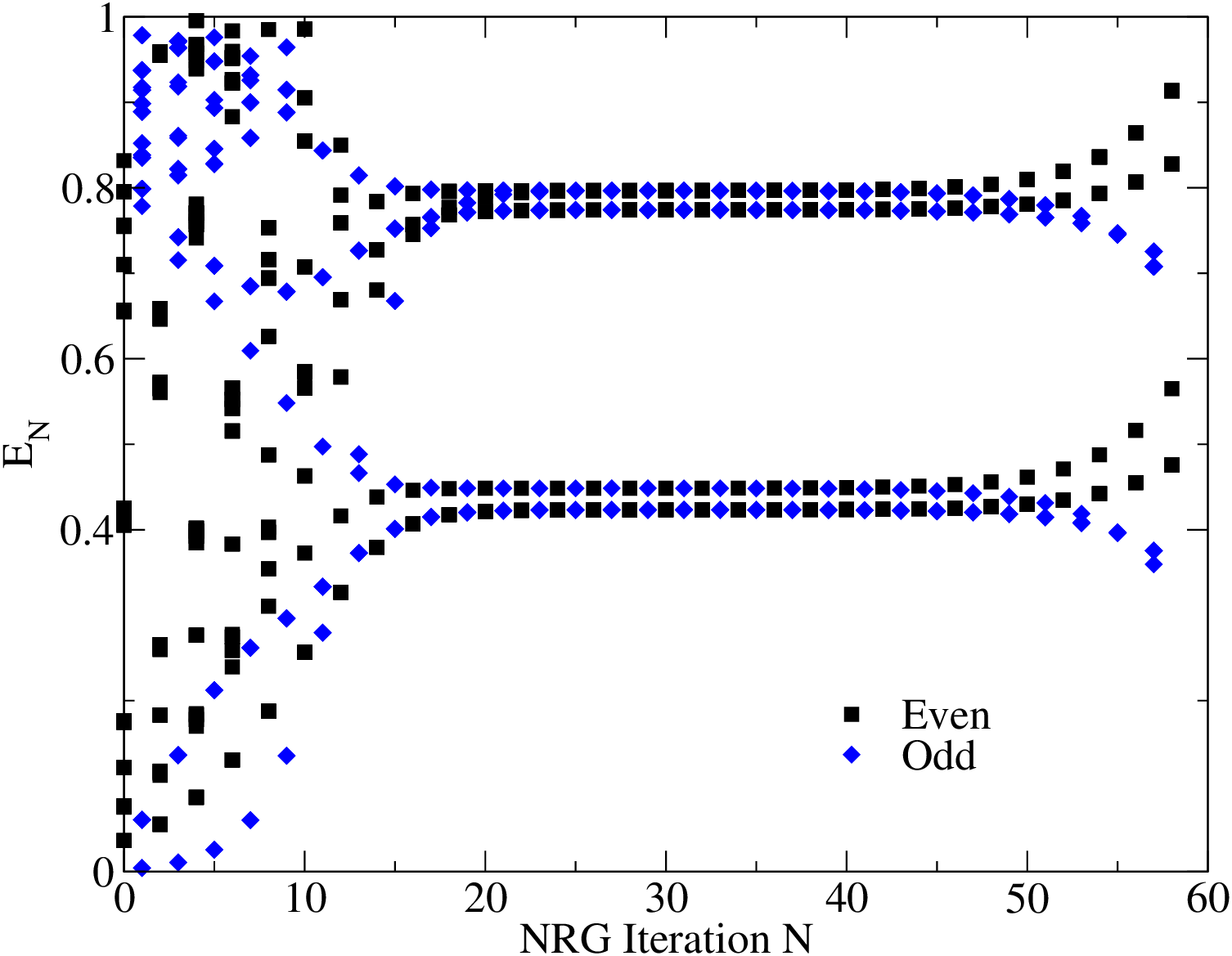}
\caption{Flow diagram of the eigenvalues obtained by NRG as a function of the iteration number 
for $D = 0.0031961 \simeq  D_c$. 
Different symbols are used for even and odd iteration numbers. The ground state energy is set to zero.
Other parameters as in the main paper.}
\label{nrg_spectra}
\end{center}
\end{figure}

The behavior of the low-energy eigenvalues of the NRG Hamiltonians $H_N$ with the iteration number $N$ gives important information 
about the fixed points of an impurity system \cite{hewson97-sm}. 
In Fig. \ref{nrg_spectra}, the NRG eigenvalues ($E_N$) of lower energy as a function of the 
length of the Wilson chain  $N$ 
are shown, for an anisotropy 
close to $D_c$ in the Anderson model (same parameters as in the main paper). At intermediate $N$ ($15 \lesssim N \lesssim 55$) there is an extended plateau signalling the 
existence of an unstable fixed point at intermediate energies or temperatures, in complete agreement with the non-Fermi 2CK fingerprints
we have found in the impurity entropy, conductance, and spectral function. To strengthen this point, two features can be seen in Fig. \ref{nrg_spectra} 
(and at higher $E_N$ not shown in the figure): 
there is no even/odd alternation in the spectra, and the energy level spacing is not uniform, both are known results for the 2CK state 
due to its non-Fermi liquid nature \cite{pang91-sm}. 

We consider as quantum numbers of the NRG eigenvalues, $Q$  --the difference of the number of electrons with respect to the ground state-- and the spin projection $S_z$. 
The low-energy eigenvalues, beginning with the ground state, have quantum numbers ($Q=0,S_z=0$), ($Q=-1, S_z=\pm 1/2$), ($Q=-2, S_z=0$), 
($Q=1, S_z= \pm 1/2$), ($Q=0, S_z=-1, 3/2$). 
We were not able to match one to one the NRG eigenvalues of our critical $S=1$ Anderson impurity model and the known eigenvalues of the
$S=1/2$ two-channel Kondo model \cite{mitchell12-sm}, probably due to the presence of marginal operators.

\section{$S=1$ Kondo impurity coupled to two conduction bands}
In order to check that our results are generic for a magnetic impurity with two orbitals hybridized with 
two conduction bands, at or close to the $n_{\rm imp} \simeq 2$ Kondo regime, we have performed a NRG 
study of the $S=1$ Kondo impurity model coupled with two conduction bands, 
\begin{equation}
 H = \sum_{\bk \nu \alpha \sigma} \varepsilon_{\bk \nu \alpha}c^\dagger_{\bk \nu \alpha \sigma}c_{\bk \nu \alpha \sigma}
+ \sum_{\nu \alpha}J_{\nu \alpha} {\bf S}_{\rm imp}\cdot {\bf s}_{\alpha \nu} + D\left(S_{\rm imp}^z\right)^2, 
\end{equation}
where, as in the Anderson model, $\nu = L, R$ refers to the left and right leads connected with the impurity, $\alpha=xz,yz$ 
refers to the two degenerate conduction electron bands, 
${\bf S}_{\rm imp}$ is the $S=1$ spin operator of the impurity, ${\bf s}_{\nu \alpha}$ is the conduction electron 
spin density at the impurity site, and $D$ is the single-ion magnetic anisotropy. 
In what follows, we consider an equal Kondo exchange coupling of the impurity to both 
leads and both conduction bands, $J_{\nu \alpha} = J$, and take 
$J=0.2$, being the half-bandwidth of the degenerate conduction bands the unit of energy ($W=1$). 
We use discretization parameters $\Lambda=2$ and 3,  keeping up to 4000 NRG states, 
and the results are $z$-averaged with $N_z=4.$ 

The calculated Kondo temperature for $D=0$ is $T^0_K = 4.2\times 10^{-5}$ ($\Lambda=3$).  
We find again a quantum phase transition between 
fully-screened Kondo and anisotropy-quenched impurity spin phases at a finite value of the anisotropy, $D_c = 1.352\times 10^{-4}$.
Notice that the relation $D/T^0_K = 3.21$ is close to the corresponding one in the Anderson model. 

\begin{figure}[ht]
\vspace{1cm}
\includegraphics[scale=0.355]{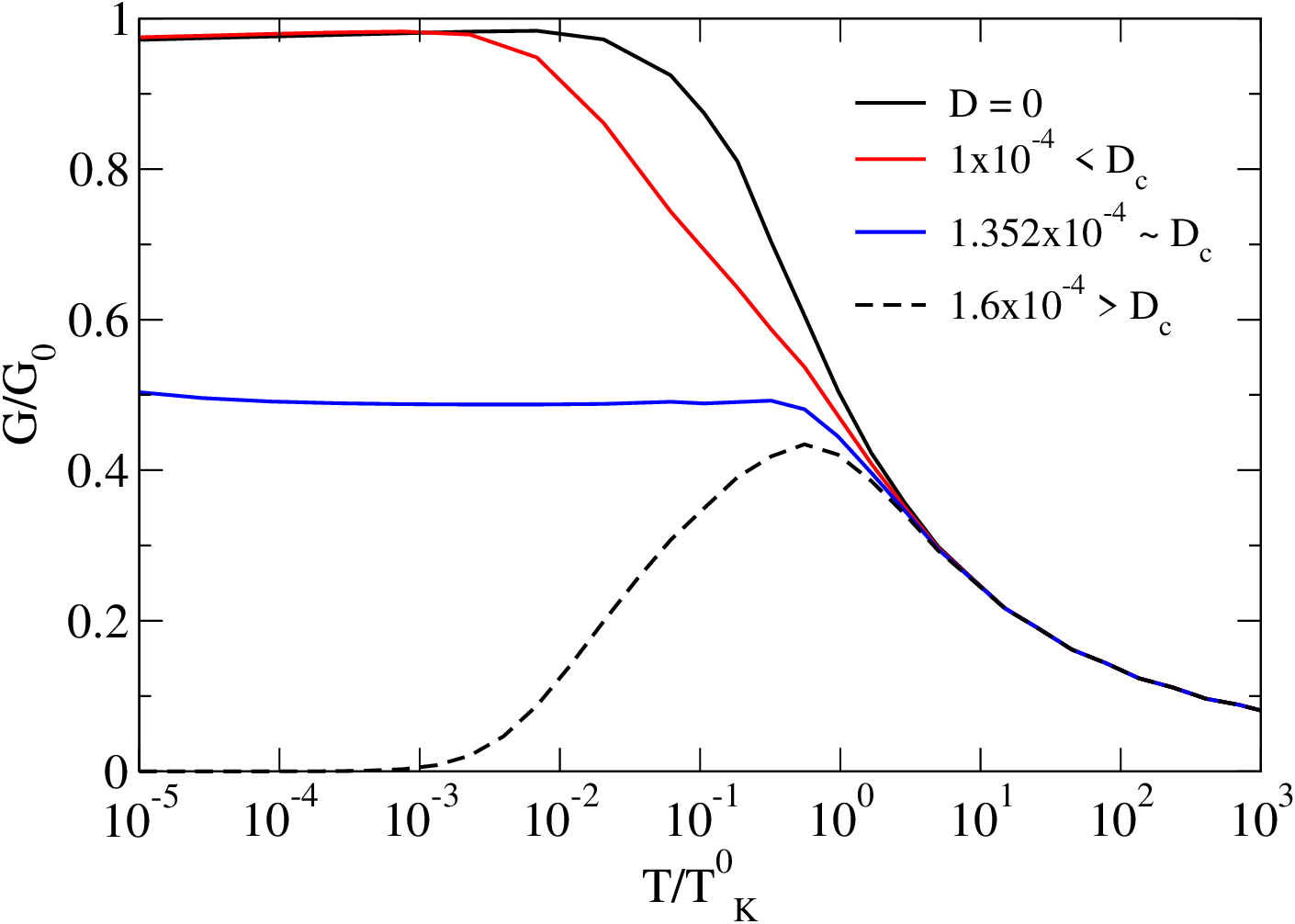}%
\begin{picture}(300,0)
\put(20,1){\includegraphics[scale=0.34]{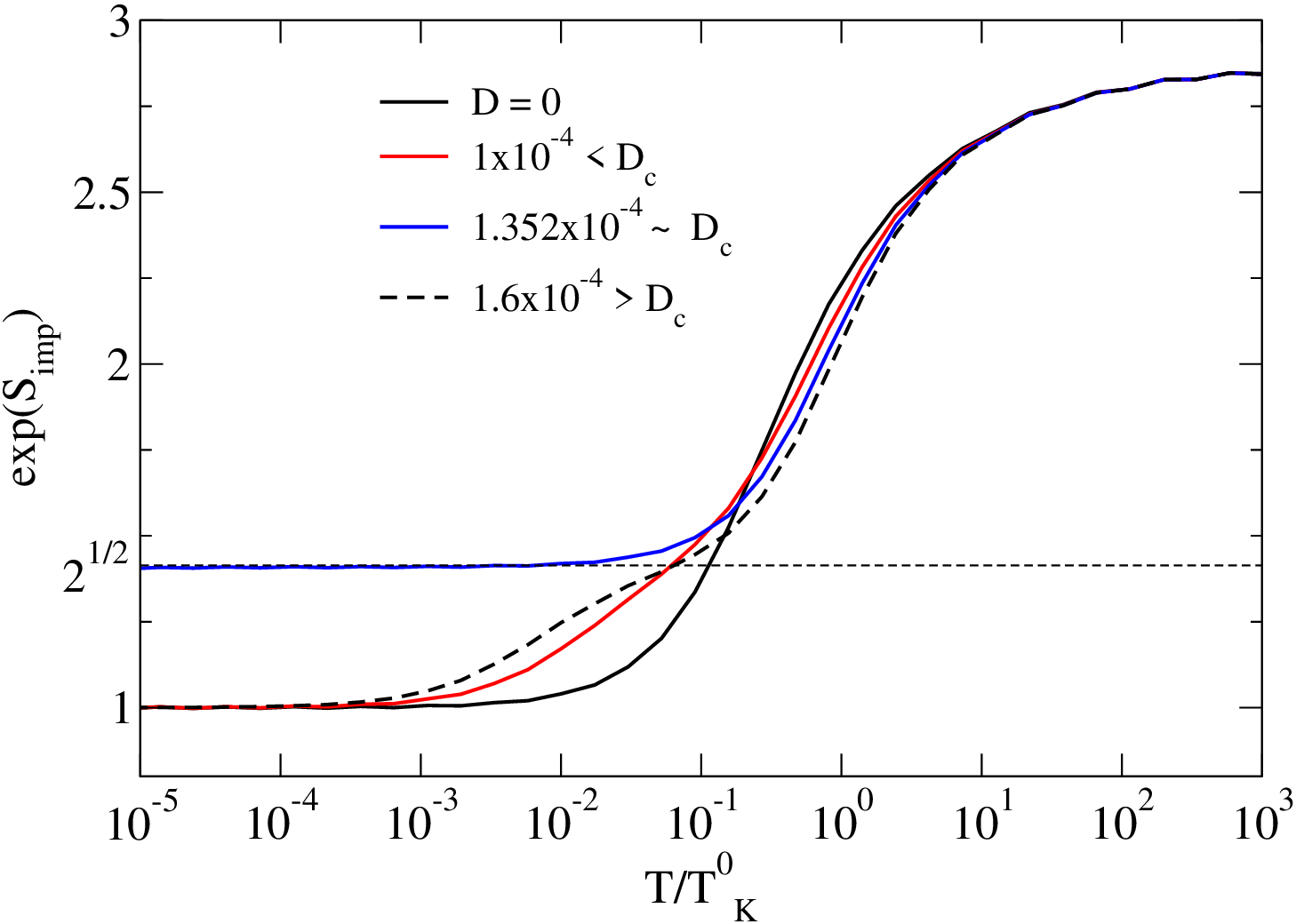}}
\end{picture}
\caption{
(Left) Electrical differential conductance as a function of $T/T^0_K$ for several values of the 
single-ion anisotropy $D.$
(Right) Impurity entropy contribution for the $S=1$ Kondo model as a function of temperature, for several values of $D$.
}
\label{SG_kondo}
\end{figure}

In the left panel of Fig. \ref{SG_kondo}, the differential conductance per channel as a function of $T/T^0_K$ is shown for different $D$ across the quantum phase transition. 
The behavior of $G$ with $D$ is the same as in the Anderson model. 
The only difference is that now the conductance (neglecting small numerical inacurracies) 
reaches the unitary limit at small temperatures
$G_\alpha(T\!\to\!0)=G_0$ for $D<D_c$, while it goes to zero
for $D > D_c$. 
This has to do with the fact that the Kondo impurity model has particle-hole symmetry. 
In the right panel of Fig. \ref{SG_kondo}, the impurity contribution to 
the entropy is plotted as a function of $T/T^0_K$ for the same anisotropies as in the conductance figure. Again, the result is qualitatively the same as for the 
Anderson model. Thus, the picture found in the main paper holds for the $S=1$ Kondo impurity: at $D_c$ there is a quantum phase transition between two Fermi liquids, and in the 
critical region there is a non-Fermi liquid phase with characteristics proper of the two-channel Kondo effect.

\begin{figure}[ht]
\begin{center}
\includegraphics*[width=0.5\columnwidth]{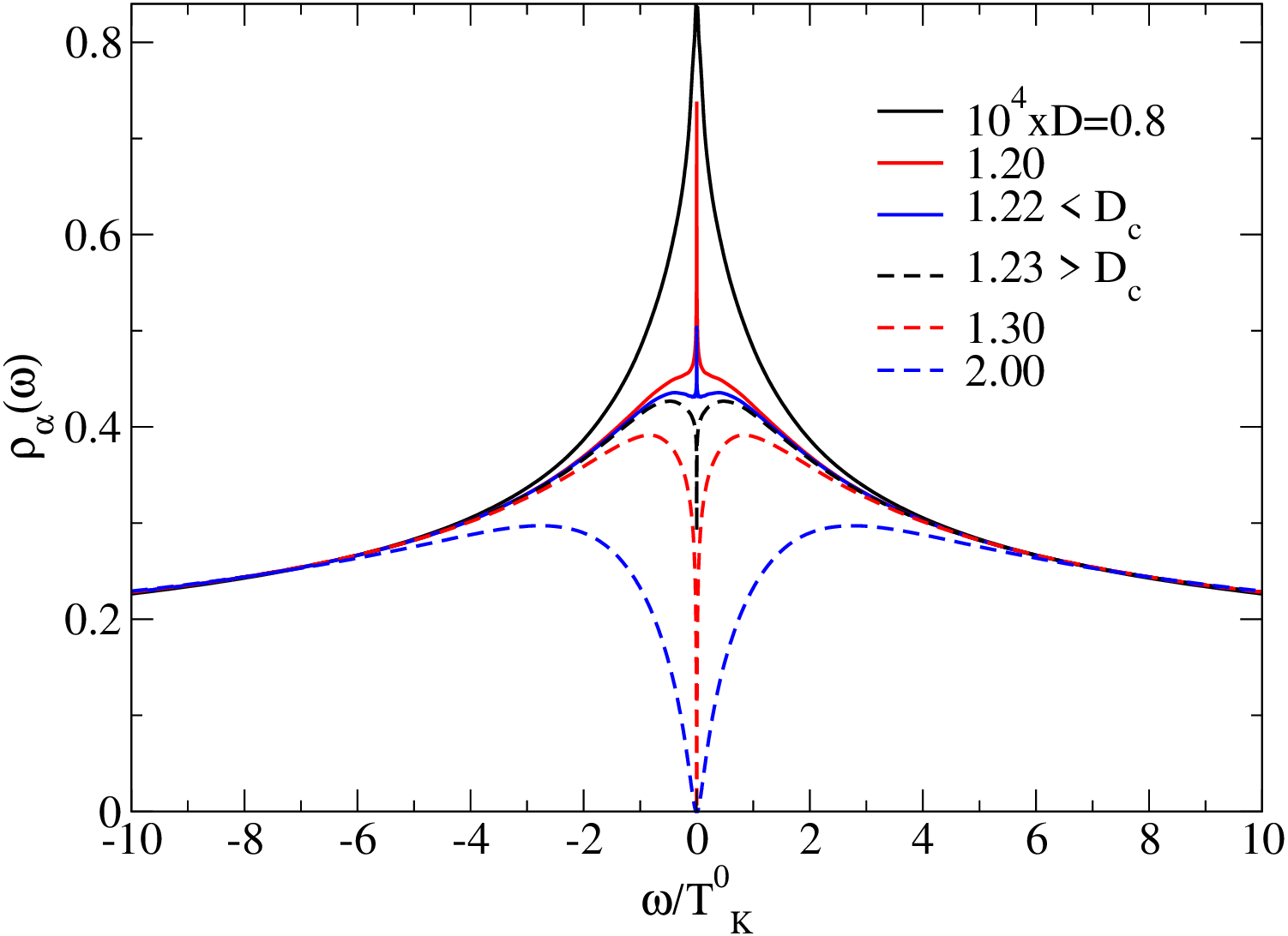}
\caption{Spectral function  $\rho_\alpha(\omega)$ versus $\omega/T^0_K$ for several values of the 
single-ion anisotropy $D$ 
for the $S=1$ Kondo impurity model.}
\label{kondo_spec}
\end{center}
\end{figure}

A function $\rho_\alpha(\omega)$ equivalent to the impurity spectral function can be defined for the Kondo model \cite{mitchell10-sm}. 
This spectral function is defined through the $t$-matrix of the conduction
electrons using the relation $\rho_\alpha(\omega) = -\pi \rho^0_\alpha \;{\rm Im}\;t_\alpha(\omega)$, where $\rho^0_\alpha=1/2W$ is the conduction electron density of states, that
is assumed energy independent. $t_{\alpha}(\omega)$ is the $t$-matrix for the conduction electrons 
with symmetry $\alpha$
scattering off the magnetic impurity, and it 
is defined by the relation 
\begin{equation}
 G_{\alpha,\bk \bk'}(\omega) = G^0_{\alpha,\bk}(\omega)\delta_{\bk \bk'} +G^0_{\alpha,\bk}(\omega) t_{\alpha}(\omega) G^0_{\alpha,\bk'}(\omega),
\end{equation}
where $G^0_{\alpha \bk}=1/(\omega + i0^+ - \varepsilon_{\bk})$ is 
the Green's function for the conduction band with symmetry $\alpha$ in the abscence of the impurity.
Fig. \ref{kondo_spec} shows $\rho_\alpha(\omega)$ as a function of $\omega/T^0_K$ for several values of $D$ \cite{note-sm}
Again, the same behavior as in the Anderson model 
is recovered. However, thanks to the particle-hole symmetry of the Kondo model around the Fermi level 
$\omega=0,$ now it is clearer that for $D \rightarrow D_c$ and $\omega \rightarrow 0$, 
$\rho_\alpha(\omega)$ takes half of its value in the fully screened Kondo phase. 
This is another characteristic of the non-Fermi-liquid two-channel Kondo effect.

\section{Topological interpretation of the Luttinger integral}

In this section, we present a topological interpretation of the Luttinger integral for the Anderson impurity model, 
following the guidelines of the recent work by Seki and Yunoki \cite{seki17-sm}.

We rewrite Hamiltonian (\ref{ham-sm}) as $H = H_0 + H_{\rm int},$ with the non-interacting part  
\begin{equation} 
H_0 = \sum_{\nu k\alpha \sigma }\varepsilon _{k}c_{\nu k\alpha \sigma }^{\dagger }c_{\nu k\alpha \sigma }+ 
\sum_{\nu k \alpha \sigma}\left( V_{\nu} {c}^\dagger_{\nu k \alpha\sigma}{d}_{\alpha \sigma} + {\rm H.c.}\right) 
+\sum_{\alpha \sigma}\tilde{\varepsilon}_d d^\dagger_{\alpha\sigma}d_{\alpha\sigma}, 
\end{equation}
where $\tilde{\varepsilon}_d$ is an effective $d$ level determined selfconsistenly, so that the total non-interacting 
occupancy $N$ is the same as the interacting one (for example, in the symmetric single-orbital Anderson model 
where ${\varepsilon}_d=-U/2$, $\tilde{\varepsilon}_d=0$). 
$H_{\rm int}=H-H_0$ includes all the two-body terms. We express the single-particle Green's function 
as an $L_s \times L_s$ matrix ${\bf G}(z),$ where $L_s$ is the number of single-particle states 
and $z$ is the complex frequency. The average electron number $N$ is given by the contour integral  
\begin{equation}
 N = \oint_\Gamma \frac{dz}{2\pi i} n_F(z) {\rm tr}\;{\bf G}(z),
 \label{Nint-sm}
\end{equation}
where the contour $\Gamma$ encloses the singularities of ${\bf G}(z)$ (all in the real-frequency axis) 
in the counterclockwise direction, as shown in Fig. 2(a) of Ref. \onlinecite{seki17-sm}, and $n_F$ is the Fermi function.
The trace can be expressed as 
\begin{equation}
{\rm tr}\;{\bf G}(z) = \sum_{k \nu \alpha \sigma}G_{k \nu \alpha \sigma}(z)+
 \sum_{\alpha\sigma}G_{d \alpha\sigma}(z),
 \end{equation}
where $G_{k\nu\alpha\sigma}$ and $G_{d\alpha\sigma}$ are the diagonal elements of ${\bf G}(z)$.
In the following, we explicitly consider that the Green's functions do not depend on the orbital and spin indices.
By the equation-of-motion method, we can obtain the relation
\begin{equation}
 G_{k\nu}(z) = \frac{1}{z-\varepsilon_k} + |V_\nu|^2 \left(\frac{1}{z-\varepsilon_k}\right)^2 G_{d}(z),
\end{equation}
what brings us to the expression
\begin{equation}
{\rm tr}\;{\bf G}(z)=  \sum_{k \nu \alpha \sigma}\frac{1}{z-\varepsilon_{k}} +\sum_{\alpha\sigma}\left(1-\frac{\partial \Gamma(z)}{\partial z}\right)  G_{d}(z),
\label{traza-sm}
\end{equation}
where $\Gamma(z) = \sum_{k \nu}|V_\nu|^2\frac{1}{z-\varepsilon_k}$ is the complex hybridization function, whose imaginary part is $\Delta$.

Using the Dyson equation,
\begin{equation}
G_{d}(z) = \frac{1}{z-\tilde{\varepsilon}_d-\Gamma(z)-\Sigma_{d}(z)},
\end{equation}
where $\Sigma_{d}(z)$ is the impurity self-energy due to $H_{\rm int}$,
(\ref{traza-sm}) can be rewritten as 
\begin{equation}
 {\rm tr}\;{\bf G}(z)=  \sum_{k \nu \alpha \sigma}\frac{1}{z-\varepsilon_{k}} +\sum_{\alpha\sigma}\frac{\partial \log (G_{d})^{-1}}{\partial z} + 
 \sum_{\alpha\sigma}\frac{\partial \Sigma_{d}(z)}{\partial z} G_{d}(z).
 \label{trace-sm}
\end{equation}

As we turn off the interactions ($H_{\rm int}=0$), the electron number is given also by the contour integral
\begin{equation}
 N = \oint_\Gamma \frac{dz}{2\pi i} n_F(z) {\rm tr}\;{\bf G}^0(z), 
 \label{N0-sm}
\end{equation}
where ${\bf G}^0(z)$ is the single-particle Green's function corresponding to $H=H_0$.
If we equal (\ref{Nint-sm}) and (\ref{N0-sm}), using the expression (\ref{trace-sm}) (and taking into account that $\Sigma_{d\alpha\sigma}=0$ for $H_{\rm int}=0$), 
we get 
\begin{equation}
 \oint_\Gamma \frac{dz}{2\pi i} n_F(z) \frac{\partial \log D_{d}(z)}{\partial z} = 
 -\oint_\Gamma \frac{dz}{2\pi i} n_F(z) \frac{\partial \Sigma_{d}(z)}{\partial z}G_{d}(z),
 \label{vollut-sm}
\end{equation}
where $D_{d}(z)=G^{0}_{d}(z)/G_{d}(z)$.
Eq. (\ref{vollut-sm}) corresponds to Eqs. (51) and (53) of Ref. \onlinecite{seki17-sm} for the Anderson impurity Hamiltonian. The contour integral in its left-hand side 
is associated with the deviation of the Luttinger volume from the non-interacting one due to many-body interactions in extended systems.

If $F(z)$ is an analytical function in the complex plane except, perhaps, on the real axis, the following relation holds:
\begin{equation}
 \oint_\Gamma \frac{dz}{2\pi i}n_F(z) F(z) = -\frac{1}{\pi}{\rm Im}\;\int_{-\infty}^{\infty}n_F(\omega)F(\omega+i\eta)d\omega, 
\end{equation}
where $\eta$ a positive infinitesimal. Using this relation and Eq. (\ref{vollut-sm}), the Luttinger integral (4) can be written as 
\begin{equation}
 I_L = -\pi \lim_{T\to 0} \oint_\Gamma \frac{dz}{2\pi i} n_F(z) \frac{\partial \Sigma_{d}(z)}{\partial z}G_{d}(z) = 
 \pi \lim_{T \to 0}\oint_\Gamma \frac{dz}{2\pi i} n_F(z) \frac{\partial \log D_{d}(z)}{\partial z}.
\end{equation}
At zero temperature, the contour integral $\Gamma$ is reduced to the contours $\Gamma_<$ and $\Gamma_0$, which enclose the negative real axis and the origin, respectively 
(see Fig. 2(b) in Ref. \onlinecite{seki17-sm}), and 
\begin{equation}
 \lim_{T \to 0}\oint_\Gamma \frac{dz}{2\pi i} n_F(z) \frac{\partial \log D_{d}(z)}{\partial z} = n_{D_d}(\Gamma_<) + \frac{1}{2}n_{D_d}(\Gamma_0),
 \label{intD-sm}
\end{equation}
where, for a contour ${\cal C}$ in the complex $z$ plane,  $n_{D_d}({\cal C})$ is the winding number that counts the (signed-)number of times that the contour $D_d({\cal C})$ 
encloses the origin of complex $D_d$ plane.

Consequently, the Luttinger integral,
$$I_L = \pi n_{D_d}(\Gamma_<)+\frac{\pi}{2}n_{D_d}(\Gamma_0),$$
has a clear topological character. In Ref. \onlinecite{seki17-sm}, the relation between these winding numbers and the
singularities (poles and zeros) of the single-particle Green’s function is stressed. 

\subsection*{Luttinger integral for the Anderson atom} 
In order to illustrate how the analytical structure of the Green's function may change with a varying 
Hamiltonian parameter, and its relation with the Luttinger integral, we analyze the evolution 
with $U$ of the Anderson impurity model in the atomic limit. 
The Hamiltonian is 
$$\hat{H}_{\rm at}=\varepsilon_d \hat{n}_d+U\hat{n}_{d\uparrow}\hat{n}_{d\downarrow}.$$
The impurity Green's function can be exactly calculated using the equation-of-motion method, and it is given by 
\begin{equation}
 G_{d\sigma}(z)=\left(1-\frac{n_d}{2}\right)\;\frac{1}{z -\varepsilon_d}+\frac{n_d}{2}\;\frac{1}{z-\left(\varepsilon_d+U\right)},
\end{equation}
where $n_d$ is the impurity occupancy. 
The Dyson equation allows us to express the Green's function in terms of the impurity self-energy
$$G_{d\sigma}(z)=\frac{1}{z-\varepsilon_d-\Sigma_{d\sigma}(z)},$$
where 
\begin{equation}
 \Sigma_{d\sigma}(z)=\frac{n_d}{2}U+\frac{\left(1-\frac{n_d}{2}\right)\frac{n_d}{2}U^2}{z-\varepsilon_d-
 \left(1-\frac{n_d}{2}\right)U}.
\end{equation}

If $\varepsilon_d < 0$ and $\varepsilon_d+U < 0$, then the electron occupancy is given by $n_d =2$ and 
\begin{equation}
 G_{d\sigma}(z) = \frac{1}{z-(\varepsilon_d+U)},\;\;\Sigma_{d\sigma}(z)=U.
\end{equation}
Trivially, the Luttinger integral vanishes in this case. 
On the other hand, we take the non-interacting Green's function $G^0_{d\sigma}(z)=1/(z-\varepsilon_d)$ (which gives the same 
interacting occupancy $n_d=1$), and, consequently,  
\begin{equation}
 D_d(z) = \frac{G^0_{d\sigma}(z)}{G_{d\sigma}(z)} = \frac{z-(\varepsilon_d+U)}{z-\varepsilon_d}. 
\end{equation}
As \begin{equation}
 \frac{\partial \log D_d(z)}{\partial z}=\frac{1}{z-(\varepsilon_d+U)}-\frac{1}{z-\varepsilon_d},
\end{equation}
the winding number of $D_d(z)$ vanishes.

For $\varepsilon_d < 0$ and $\varepsilon_d+U > 0$, the electron occupancy is $n_d=1,$ yielding
\begin{equation}
 G_{d\sigma}(z) = \frac{1}{2}\left[\frac{1}{z-\varepsilon_d} + \frac{1}{z-(\varepsilon_d+U)}\right],\;\;\;\Sigma_{d \sigma}(z)= 
 \frac{U}{2}+\left(\frac{U}{2}\right)^2\frac{1}{z-(\varepsilon_d+\frac{U}{2})}. 
\end{equation}
Notice that, in this case,  the Green's function adquires simultaneously a new pole ($z_p=\varepsilon_d$) and a zero 
($z_o=\varepsilon_d+U/2$). The last one can be seen as a singularity of the self-energy.

As 
\begin{equation}
 G_{d\sigma}(z)\frac{\partial \Sigma_{d\sigma}(z)}{\partial z}=
\frac{1}{z-\left(\varepsilon_d+\frac{U}{2}\right)}-\frac{1}{2}\left[\frac{1}{z-\varepsilon_d}+\frac{1}{z-\left(\varepsilon_d+U\right)}\right],
\end{equation}
the Luttinger integral $I_L = -\pi/2$ ($+\pi/2$) if $\varepsilon_d$ is below (above) the symmetric point, $\varepsilon_d= -U/2$.
In order to get a non-interacting limit with the same occupancy, $n_d=1$, we need to shift $\varepsilon_d$ to the Fermi level ($\tilde{\varepsilon}_d=0$), 
so $G^0_{d\sigma}(z)=1/z.$
As 
\begin{equation}
 D_d(z) = \frac{G^0_{d\sigma}(z)}{G_{d\sigma}(z)} = \frac{(z-\varepsilon_d)(z-(\varepsilon_d+U))}{z(z-(\varepsilon_d+\frac{U}{2}))},
\end{equation}
the countour integral (\ref{intD-sm}) is $-1/2$ ($+1/2$) if $\varepsilon_d$ is below (above) the symmetric point, in agreement with the Luttinger integral 
values given above.


\begin{thebibliography}{99}
\bibitem{note} The integral $I_L$ is in Eq. (63) of Ref. \onlinecite{luttinger60b}, Eq. (23) 
of Ref. \onlinecite{langreth66}, and Eq. (3.3) of Ref. \onlinecite{yoshimori82}. 
In our case it is given by Eq. (\ref{il}).
\end{thebibliography}
\begin{thebibliography}{99}
\bibitem{note-sm} For the computation of the spectral function $\rho_\alpha$, we take $\Lambda=2$ and we keep 3000 NRG states. 
As a consequence, $D_c = 1.225\times10^{-4}$ and $T^0_{K}=3.8\times 10^{-5}$ are a little different from the corresponding 
values that appear in the conductance and impurity entropy in Fig. \ref{SG_kondo} ($D_c=1.352\times 10^{-4}$,  $T_K^0 = 4.2\times 10^{-5}$).
\end{thebibliography}
\end{document}